\documentclass{sf2a-conf2017}
\usepackage{graphicx}
\usepackage{hyperref}
\usepackage[]{natbib}  
\usepackage{epstopdf}

\def\BibTeX{{\rm B\kern-.05em{\sc i\kern-.025em b}\kern-.08em
    T\kern-.1667em\lower.7ex\hbox{E}\kern-.125emX}}
\bibpunct{(}{)}{;}{a}{}{,}  

\begin{document}

\TitreGlobal{SF2A 2017}


\title{On the derivation of radial velocities of SB2 components: a ``CCF vs TODCOR'' comparison$^*$\footnote{* based on observations performed at the
Haute-Provence Observatory}}

\runningtitle{RVs of SB2 components}

\author{J.-L. Halbwachs}\address{Universit\'e de Strasbourg, CNRS, Observatoire astronomique de Strasbourg, UMR 7550, F-67000 Strasbourg, France }

\author{F. Kiefer}\address{Institut d'Astrophysique de Paris, CNRS, UMR 7095, 98bis boulevard Arago, F-75014 Paris}

\author{F. Arenou}\address{GEPI, Observatoire de Paris, PSL Research University, CNRS, Universit\'e Paris Diderot, Sorbonne Paris Cit\'e, Place Jules Janssen, F-92195 Meudon, France}

\author{B. Famaey$^1$}

\author{P. Guillout$^1$}

\author{R. Ibata$^1$}

\author{T. Mazeh}\address{School of Physics and Astronomy, Tel Aviv University, Tel Aviv 69978, Israel}

\author{D. Pourbaix}\address{FNRS, Institut d'Astronomie et d'Astrophysique, Universit\'{e} Libre de Bruxelles, boulevard du Triomphe, 1050 Bruxelles, Belgium}




\setcounter{page}{237}


\maketitle


\begin{abstract}
The radial velocity (RV) of a single star is easily obtained from cross-correlation of the spectrum with a template, but the treatment of double-lined spectroscopic binaries (SB2s) is more difficult. Two different approaches were applied to a set of SB2s: the fit of the cross-correlation function with two normal distributions, and the cross-correlation with two templates, derived with the TODCOR code. It appears that the minimum masses obtained through the two methods are sometimes rather different, although their estimated uncertainties are roughly equal. Moreover, both methods induce a shift in the zero point of the secondary RVs, but it is less pronounced for TODCOR. All-in-all the comparison between the two methods is in favour of TODCOR.
\end{abstract}



\section{Introduction}
The derivation of a radial velocity (RV) from a CCD spectrum is a routine operation for a single-lined spectroscopic binary (SB1),
leading to an accuracy of a few m.s$^-1$, or even less. However, things are not so simple when double-lined binaries (SB2s)
are considered. However SB2s allow the estimation of the masses of the stellar component when the orbital inclination may be obtained
from an astrometric technique, such as interferometry or high-precision spatial astrometry, and accurate RVs are necessary
to derive accurate masses. For that reason, we have applied the two most common techniques on a set of well-observed SB2s,
and we compare their results hereafter.

\section{The SB2 sample}
We consider 24 SB2s which were observed since 2010 with the SOPHIE spectrograph installed on the 193 cm telescope of the
Haute-Provence Observatory (OHP). These stars are all known spectroscopic binaries for which it could be possible to 
derive the masses with an accuracy around 1~\% when the astrometric measurements of the Gaia satellite will be delivered
\citep{paper1}. Ten revised spectrocopic orbits were published in \cite{paper3}, and the publication of 14 others is in
preparation \citep{paper4}. These orbits are presented in \cite{sf2a17b}


\section{The ``CCF1'' technique}
\begin{figure}[ht!]
 \centering
 \includegraphics[width=0.8\textwidth,clip]{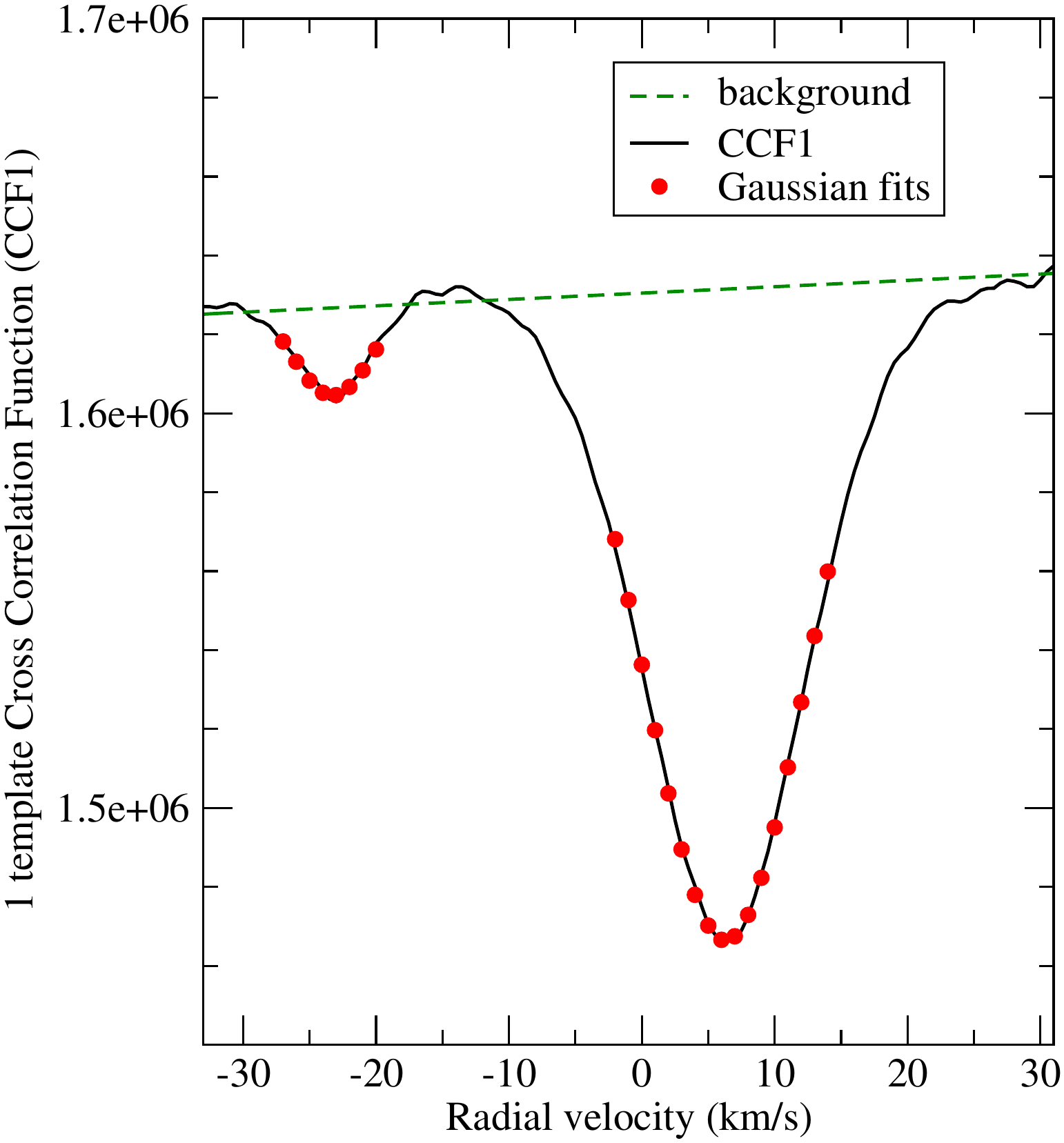}      
  \caption{The cross-correlation function of a spectrum of a SB2 with a template. The red points are the sum of two
normal functions derived from a range of 1.4~$\sigma$ around  the RV of the components.}
  \label{Halbwachs1:fig1}
\end{figure}

The first technique we have applied makes use of the cross-correlation functions (CCFs) of the spectra with a single template. 
In practice, it is the numerical equivalent of the photoelectric RVs measured, for instance, with the now decommissionned
CORAVEL instrument \citep{Baranne79}: each component generates in the CCF a bell-shaped dip which is assumed to obey a
normal function. Therefore, the velocities are derived by computing the parameters of two normal functions which the sum
is as close as possible to the CCF. For any SB2, the slope of the background is fixed to the same value for all the spectra;
at the opposite, the middles, the standard deviations, and the depths of the correlation dips are calculated from a $\chi^2$ 
minimization for the CCF of each spectrum. 
In practice, these parameters are derived from a range
restricted to a given number of standard deviations, $\sigma$, around each minimum as shown in Fig.~\ref{Halbwachs1:fig1}. This range
may be as large as 2 or even 3~$\sigma$ for some stars, but, for others, even 1~$\sigma$ would lead to a very approximative fit;
this depends on the compatibility of the templates with the two spectra, and also on the rotation velocities of the stars.
As a consequence, we may expect that the RVs of the components are affected by systematic errors when the correlation dips
are closer than approximately 3 times the sum of their standard deviations. However, reality is often even worse: since
the template doesn't exactly correspond to the spectrum of each component, the correlation dips are often flanked by side lobes
which are much less deep but as large as the main dips. Therefore, the minimum difference guaranteeing reliable RVs
is in fact 6 times the sum of the standard deviations.
Since the standard deviations are as large as about 3 km/s
for the slow rotators with G-K spectral types, and larger otherwise, that means that, in the best case, the RVs are dubious when the difference
is less than 36 km/s; this concerns a large part of the measurements obtained for our sample.

\section{TODCOR}
The TODCOR alrgorithm \citep{TODCOR,TODCOR2} derives the CCF assuming a template for each component. This method is rather
sophisticated, since the templates must be chosen carefully, as explained in \cite{paper3}. For the search
of the minimum of the CCF, see, e.g., \cite{sf2a2013}.
The templates are synthetic spectra extracted from the Phoenix library \citep{Hauschildt99}. When they are really
similar to the actual spectra of the components, the errors related to the CCF1 methods
are avoided. However, when the templates are different, systematic errors related to the difference of
RV may rise again.

The RV used to derive the published orbits were all obtained with
TODCOR.

\section{A comparison CCF1 vs TODCOR}
Rather than comparing the RVs coming from the two techniques, we used them to derive the SB2 orbital elements,
following the method presented in \cite{paper3}. A systematic shift of the RVs of the secondary components is added
to the classical orbital elements, in order to verify the adequacy of the templates. In this section, the results
of both methods are compared, considering three points : the minimum masses of the components, 
the shift of the secondary RVs, and the standard deviations of the residuals of the orbits.

\subsection{Minimum masses}

\begin{figure}[ht!]
 \centering
 \includegraphics[width=0.8\textwidth,clip]{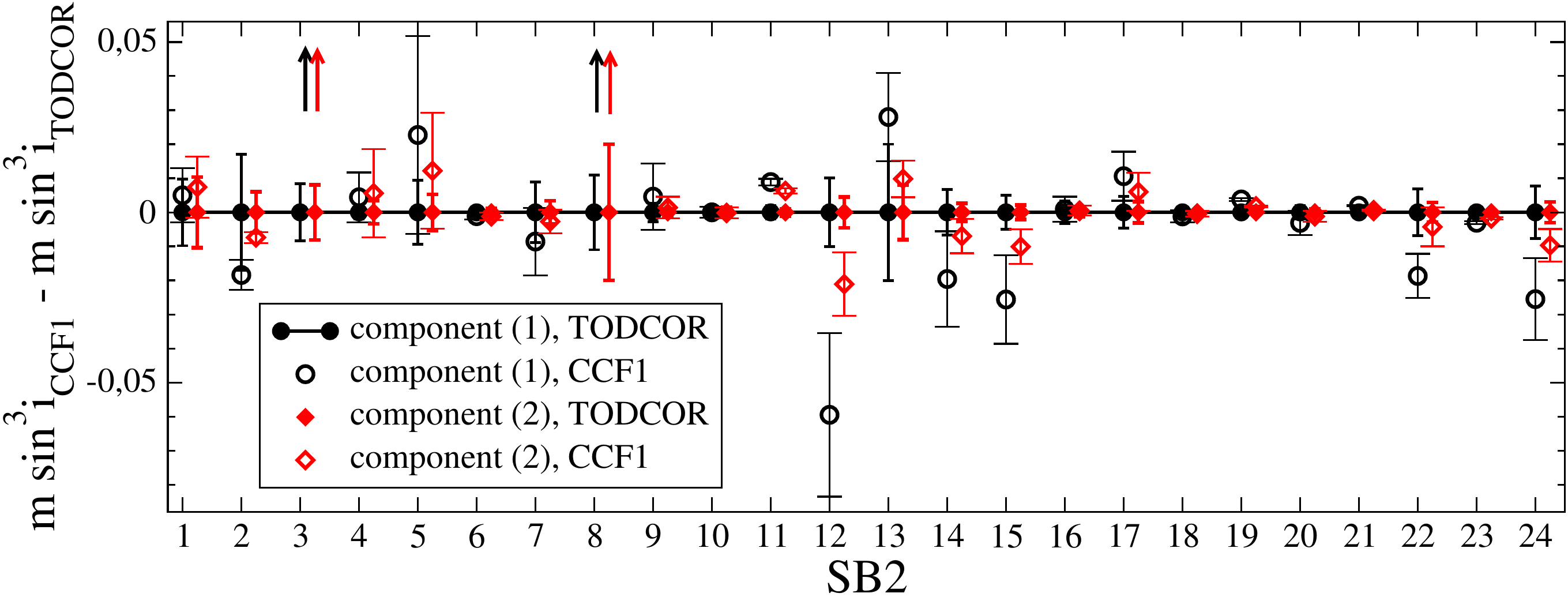}      
  \caption{Comparison of the minimum masses, ${\cal M}_1 \sin^3 i$ and ${\cal M}_2 \sin^3 i$, derived from RVs obtained
with the CCF1 and with the TODCOR method. The masses coming from TODCOR are used as references. The stars are sorted according to
the spectral type of the primary component, with the early-type stars at left.}
  \label{Halbwachs1:fig2}
\end{figure}

The orbital elements of a SB2 orbit lead to the minimum masses, ${\cal M}_1 \sin^3 i$ and ${\cal M}_2 \sin^3 i$, where $i$ is the
inclination of the orbital plane and ${\cal M}_1$ and ${\cal M}_2$ the masses of the components. The minimum masses were derived
from the RVs obtained with the CCF1 and with the TODCOR methods, and they are compared in Fig.~\ref{Halbwachs1:fig2}. The error
bars are represented too.

It appears that the difference has the same sign for both components, and, more important, that the minimum masses may be 
very significantly different. This confirms the importance of the choice of the technique.
.
\subsection{Shift of the secondary RVs}

\begin{figure}[ht!]
 \centering
 \includegraphics[width=0.8\textwidth,clip]{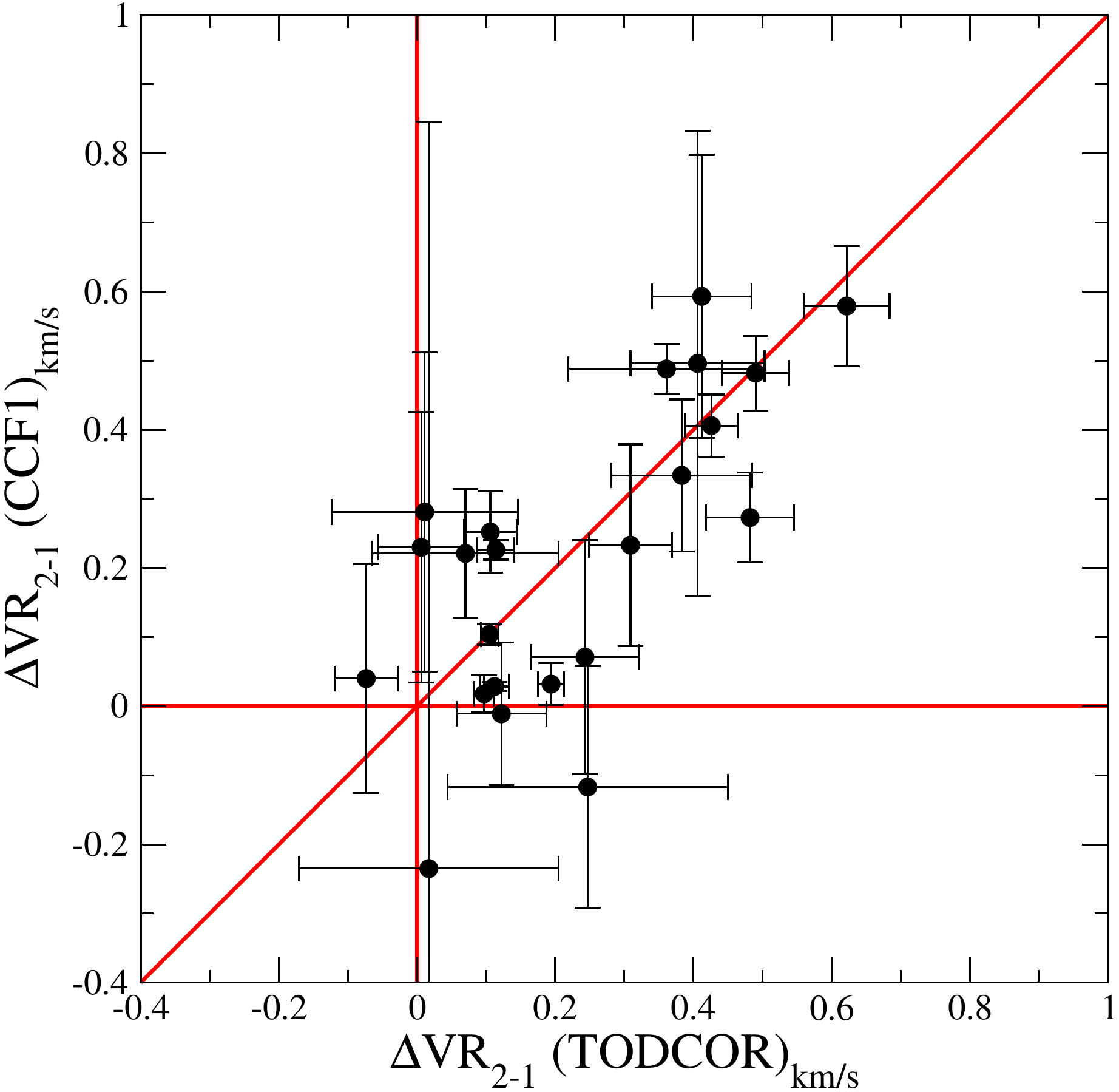}      
  \caption{Comparison of the differences between the secondary and the primary mean RVs.
The points above the diagonal represent the SB2s for which the CCF1 method gives a difference larger than
that obtained with TODCOR.}
  \label{Halbwachs1:fig3}
\end{figure}

The shift of the secondary RVs with respect to the primary ones is presented in Fig.~\ref{Halbwachs1:fig3}.
Aside from a few exception, this difference is always positive. For the CCF1
method this comes obviously from the choice of the template which has a spectral type earliest that the
secondary component.We notice also a concentration of stars along the vertical red line: these stars
have a negligible shift when the TODCOR method is applied, as expected. However, we see also a lot
of SB2s along the red diagonal, and for which the shift is roughly the same with TODCOR than with CCF1.

\subsection{Residuals of the orbit}

\begin{figure}[ht!]
 \centering
 \includegraphics[width=0.8\textwidth,clip]{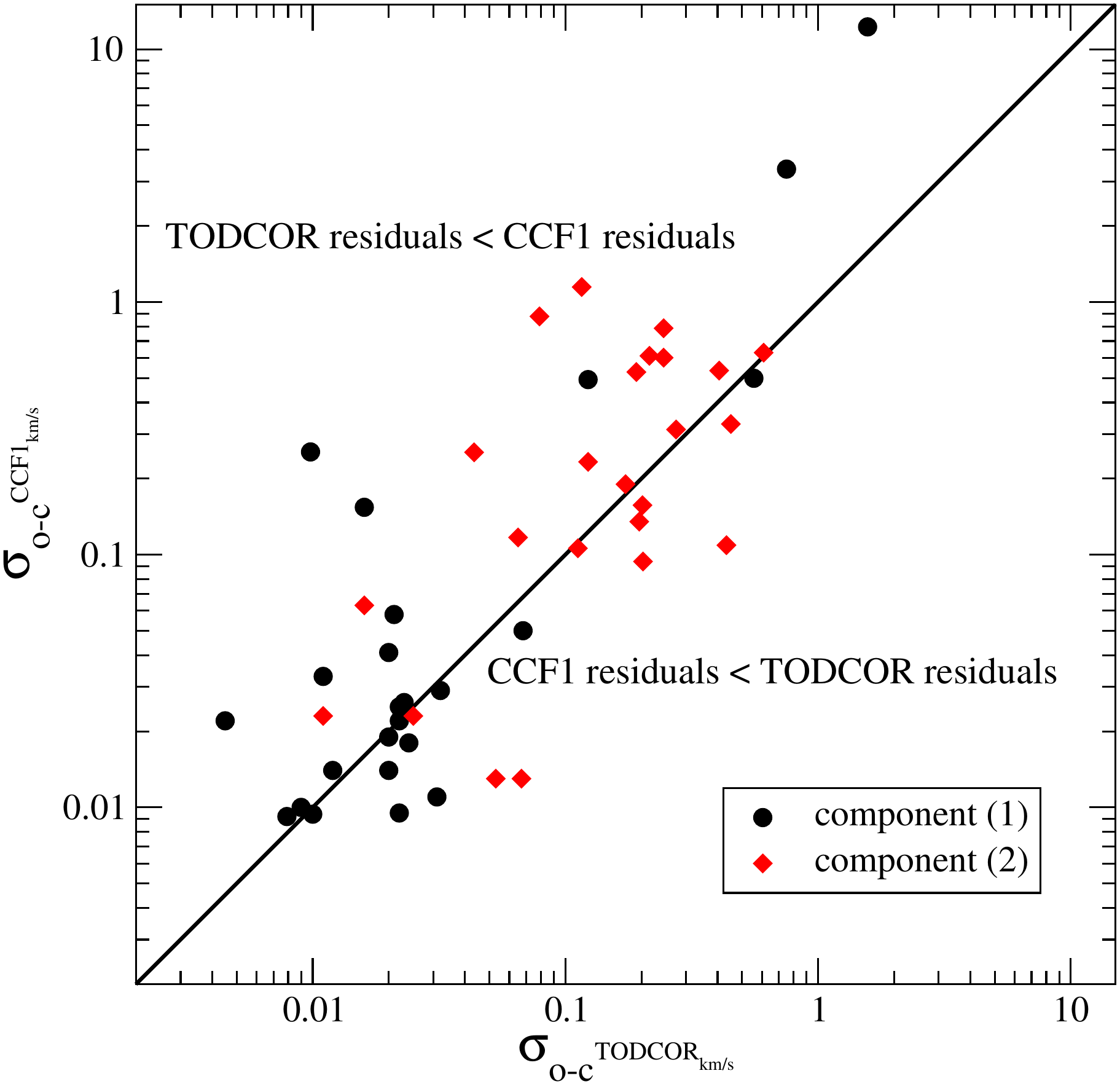}      
  \caption{Comparison of the standard deviations of the residuals of the RVs after the calculation of the
SB2 orbital elements.
The points above the diagonal represent the SB2s for which the CCF1 method gives residuals larger than
that obtained with TODCOR.}
  \label{Halbwachs1:fig4}
\end{figure}

The residuals of the SB2 orbits are presented in Fig.~\ref{Halbwachs1:fig4}. The stars are concentrated along the
diagonal, but an excess of large residuals is visible for the CCF1 method.

\section{Conclusions}

We have seen that TODCOR and the CCF1 method give RVs which are clearly different, since the minimum masses
derived from the orbital SB2 elements are often not compatible. It is expected that TODCOR give more
reliable results than CCF1, and it is confirmed that the smallest systematic shift between the RVs of the components
is obtained with TODCOR, on average. Moreover, TODCOR leads to residuals which are, on average, smaller than
those obtained from CCF1. Therefore, we confirm that TODCOR is the most reliable technique. Nevertheless, it
is not perfect: the shift of the secondary RVs is not systematically negligible, as it should be, and the
residuals of TODCOR are sometimes larger than that of CCF1. This is probably due to differences between the
actual spectra and the templates from the Phoenix library.

\begin{acknowledgements}
This project was supported by the french INSU-CNRS ``Programme National de Physique Stellaire''
and ``Action Sp\'{e}cifique {\it Gaia}''. We are grateful to the staff of the
Haute--Provence Observatory, and especially to Dr F. Bouchy, Dr H. Le Coroller, Dr M. V\'{e}ron, and the night assistants, for their
kind assistance. We made use of the SIMBAD database, operated at CDS, Strasbourg, France. This research has received funding from the European Community's Seventh Framework Programme (FP7/2007-2013) under grant-agreement numbers 291352 (ERC)
\end{acknowledgements}



%
\end{document}